# Stabilization of mechanical strength in a nanocrystalline CoCrNi concentrated alloy by nitrogen alloying

Igor Moravcik[a], Markus Alfreider[a], Stefan Wurster[b], Lukas Schretter[a], Antonin Zadera[c], Vítezslav Pernica[c], Libor Čamek[c], Jürgen Eckert[a, b], Anton Hohenwarter[a]

a - Chair of Materials Physics, Department of Materials Science, Montanuniversität Leoben, Jahnstrasse 12, 8700 Leoben, Austria.

b - Erich Schmid Institute of Materials Science, Austrian Academy of Sciences, Jahnstrasse 12, 8700 Leoben, Austria.

c - Institute of Manufacturing Technology, Brno University of Technology, Technicka 2896/2, 61669 Brno, Czech Republic

Corresponding author: Igor Moravcik, E-mail: igor.moravcik@unileoben.ac.at; moravcik.tk@gmail.com

Tel: +421 911 566 030; https://orcid.org/0000-0001-6034-8779

***Abstract:*** *The mechanical performance and microstructures of a CoCrNi medium-entropy alloy (MEA) and NCoCrNi, alloyed with 0.5at% N, after high pressure torsion (HPT) and subsequent annealing treatments in a temperature range of 150-1000 °C were investigated. The introduction of N results in a reduction of grain size by 40% and ~10-15% increased hardness and bending strength after HPT. Both materials showed strain hardening and plasticity after HPT even at these strength levels over 1500 MPa which is induced by the saturated nanocrystalline state. Annealing at intermediate temperatures of 300-500 °C resulted in an additional increase of hardness and strength by ~20-40%, compared to the HPT deformed state. This effect was ~10 % more pronounced in the nitrogen alloyed CoCrNi that also showed better thermal stability. However, the increase in strength after 500 °C annealing was accompanied by a drastic loss of ductility in both materials.*

## 1. Introduction

Medium entropy alloys (MEAs), sometimes also referred to as concentrated solid solution alloys (CSSA), have obtained substantial interest of the scientific community over the last years [1–3]. Compared to traditional alloys that are mostly based on alloying a single, major element, the trademark feature of MEAs is the lack of a base element. Instead, MEAs are composed of 3-4 elements in equal or near-equal atomic concentrations [4] leading to increased compositional freedom while, at the same time, retaining a simple solid solution microstructure. Some MEAs are characterized by enhanced combinations of room temperature strength and ductility [5], as well as high temperature strength [3,6]. Due to the relatively high concentrations of passivating elements such as Cr, several MEA further exhibit high oxidation and corrosion resistance [7,8]. The most Cr-rich MEAs show better corrosion properties compared to 300-series stainless steels [9][10] or even some Ni-base alloys [10].

Presently, the equimolar CoCrNi alloy with a single-phase face centered cubic (FCC) microstructure is one of the most interesting MEAs. The mechanical and most functional properties of this material surpass those of the intensively studied Cantor CoCrFeMnNi alloy basically in every aspect [11]. In fact, the CoCrNi has shown superior mechanical properties also among other FCC MEA alloys [12], owing to its excellent combination of yield strength, tensile strength, and ductility. The fracture toughness value in terms of $K_{Ic}$ reaches 208 MPa·m$^{1/2}$ at 293 K [13], rendering CoCrNi one of the most fracture-resistant materials of today [13]. In addition to the excellent strength-ductility synergy, this alloy also shows good corrosion [9] as well as excellent high temperature oxidation resistance [13,14] due to formation of a protective $Cr_2O_3$ layer [15][16]. Importantly, the alloy also shows enhanced resistance to hydrogen embrittlement [17]. The CoCrNi alloy can be efficiently produced by traditional as well as by powder metallurgy routes [13,18].

Recently it was shown that the yield strength of the base CoCrNi material can be substantially increased by nitrogen alloying without deteriorating ductility [19]. This was related to an increase of the lattice friction and Hall-Petch coefficient by the introduction of nitrogen [20] thus increasing the stress required for the onset of dislocation plasticity without restricting the number of available slip systems.

Another strategy to further increase the strength of the material is provided by severe plastic deformation. Several studies have shown that the minimum grain size can be tailored by alloying where already small additions can drastically reduce the grain size [21], improve the mechanical strength [20] and thermal stability [22]. In addition, it was shown that the thermal stabilization of nanocystalline structures is also significant for the "hardening by annealing" phenomenon that leads to an increase of hardness instead of a softening as would normally be expected from traditionally cold-worked materials, upon thermal recovery [23][24]. The above-mentioned coarse-grained N-alloyed CoCrNi-alloy represents an uninvestigated candidate for further mechanical improvement by applying severe plastic deformation (SPD)-methods [25].

In this contribution, we study the effect of N-alloying on the microstructure and mechanical properties of the CoCrNi MEA by nano-structuring using high pressure torsion (HPT). Isochronal and isothermal annealing treatments were performed to assess the potential of the alloy for hardening by annealing and to explore the thermal stability and its kinetics. The resulting changes in mechanical properties were investigated by common Vickers hardness testing as well as in-situ micro-cantilever bending experiments to study the deformation characteristics in detail. In addition, meso-scale 3-point bending tests were performed to allow for characterization of bulk properties and fracture-surface analysis.

## 2. Experimental methods

### 2.1 Material preparation

Trapezoidal ingots of 3-element CoCrNi ($Co_{33}Cr_{33}Ni_{33}$) and NCoCrNi ($N_{0.5}Co_{33.2}Cr_{33.2}Ni_{33.2}$) alloys with shapes ~ 60 mm × 60mm × 250 mm$^3$ were produced by vacuum induction melting and casting into cast-iron molds. The melting was performed in $ZrSiO_4$ crucibles in a Heraeus LS2vacuum induction furnace, with a minimal pressure of 4 Pa during melting and casting. The ingots were homogenized at 1250 °C for 2 h in a Heraeus K 1252 atmospheric furnace and rapidly quenched in water. Foundry chunks of pure Co, Cr and Ni elements were used as raw materials. Nitrided ferrochromium with ~15 at. % of N (3.7 wt. %) was used in the NCoCrNi material as a source of N. The concentration of N in the homogenized NCoCrNi material was measured to be 0.5 at % (0.12 wt. %) by using a LECO TCH600 spectrometer, while 1 at. % (0.25 wt. %) of N was introduced to the NCoCrNi melt. The decrease in the final concentration of the ingots can be attributed to nitrogen evaporation upon melting in vacuum environment. The use of ferrochromium as N source also caused ~2 at. % of Fe in the

NCoCrNi material. Such relatively low concentration of Fe can be neglected due to similarities in Fe, Co and Ni atoms, as they are also neglected up to 2-5 wt. % concentrations in many Ni-base alloys such as Alloy 625 or WASPALOY [26].

Disks with 8 mm diameter and 1 mm thickness were cut from the ingots by electro discharge machining and processed by quasi-constrained HPT at 7.8 GPa with a rotational speed of 0.2 rotations/min at room-temperature. Further details about the method can be found elsewhere [27,28]. Five HPT rotations were used to produce the sample since comparable studies [29] have shown that with this degree of deformation a sufficient degree of homogeneity in the hardness distribution within the HPT-disks can be achieved. The specified procedure results in a cumulative shear strain γ>50 at radial positions > 1.3 mm from the center of the cylinder and induces homogenous microstructure as described in previous work [29]. All microstructural features and mechanical properties were probed in these regions.

Isochronal heat treatments were performed for 1 h at temperatures between 100°C to 1000°C using the standard atmospheric furnace (Heraeus K 1252). Additional isothermal annealing at 300°C and 500°C was carried out for times between 10 min and 150 h. Additionally, isothermal heat treatments at 500°C were performed for up to 200 h.

2.2 **Mechanical characterization**

Microhardness testing was carried out on a Zwick/Roell DuraScan device with a load of 0.98N - HV0.1 and a dwell time of 12s. Hardness center-to-circumference line measurements were firstly made on the circular surface of the HPT-processed samples similar to [27] (sketch in Fig. 1 a)), using a spacing of 0.25 mm. Hardness center-to-circumference line measurements were also performed on the polished cross-sections of HPT (sketch in Fig. 1 b)) disks that were cut in half. The rest of the hardness measurements shown in section 3.2 comparing heat-treated conditions, were made in homogenous regions on flat surfaces of the disks with radial positions > 1.3 mm, as explained before. A focused ion beam (FIB) system (Auriga, Zeiss) with $Ga^+$ ion source was used to prepare micro-cantilever bending-test samples with cross-sectional dimensions of width $W$=5 µm (dimension parallel to vector of the applied force), thickness $B$ = 5 µm (dimension perpendicular to cantilever side face) and bending length $L_b$ of ~25 µm (dimension from root to indenter tip). To remove FIB-induced artifacts such as curtaining and excessive taper angle, final FIB polishing was performed using 30 keV and 500 pA and ±1.5° degree pre-tilt, respectively. Each micro-cantilever size and loading length was measured separately from SEM images before the test, while a minimum of 3 cantilevers were tested for one material condition.
Micro-mechanical cantilever bending experiments were performed in-situ in an SEM (DSM982, Zeiss) using a microindenter system (UNAT-SEM 1, ASMEC) with a maximum applicable force of 300 mN and a wedge-shaped conductive diamond tip (Synton-MDP). The tests were performed in open loop displacement control with a loading rate of 20 nm·s$^{-1}$. The loading force $F$ and displacement $d$ were recorded at a frequency of 32 Hz. To correlate SEM observations with measured mechanical data, micrographs were captured sequentially at a frame rate of approximately 1 s$^{-1}$. The elastic bending stress $\sigma_b$, bending strain $\varepsilon_b$ pertaining to maximum stress and strain at the outermost surface layer of the cantilever, and bending modulus $E_b$ were calculated as [30]:

$$\sigma_b = 6 \frac{F \times L_b}{B \times W^2} \qquad (1),$$

$$\varepsilon_b = 3 \frac{W \times d}{2 \times L_b^2} \qquad (2),$$

$$E_b = \frac{F \times (L_b)^3}{3 \times d \times (1/12 \times B \times (W^3))} \tag{3}$$

In order to probe larger material volumes, 3-point bending experiments were performed at room temperature in a bending test module (Model 5000 N, Kammrath&Weiss) using a 2 kN load cell and a crosshead movement speed of 2.5 µm s$^{-1}$. The 3-point bending test samples had dimensions of width $W$=1 mm, thickness $B$=0.55-0.62 mm (dimension perpendicular to beam side face) and loading support span $L_{3pb}$ of 4.46 mm. The 3-point bending stress σ$_{3pb}$ and bending strain (eps) were calculated using [30]:

$$\sigma_{3pb} = 3/2 \frac{F \times L_{3pb}}{B \times W^2} \tag{4}$$

$$\varepsilon_b = 6 \frac{W \times d}{L_b^2} \tag{5}$$

### 2.3 Microstructural characterization

A Jeol 2200 FS (JEOL, Ltd.) transmission electron microscope (TEM) was used for the characterization of the samples manufactured by mechanical thinning and precision ion polishing using Ar-gas. The chemical composition of the phases was evaluated using energy-dispersive X-ray spectroscopy (EDS) using Bruker XFlash® and Oxford INCA Energy TEM 200 EDS detectors mounted on a TESCAN MAGNA scanning electron microscope (SEM) and the Jeol 2200 FS TEM, respectively. The TESCAN MAGNA SEM was also utilized for microstructural investigation with secondary electron (SE) and backscattered electrons (BSE) detectors. The electron-transparent TEM samples were also imaged using a MAGNA scanning transmission electron (STEM) detector, to gain an overview of the microstructure at intermediate magnifications, as compared to TEM. Electron backscatter diffraction (EBSD) and transmission Kikuchi diffraction (TKD) was performed using a Bruker eFlash detector. The mean grain size was estimated from EBSD scans following ISO 13067:2011 [31] using a minimum grain boundary misorientation angle of 5° and always excluding twin boundaries using the ATEX software package [32].

## 3. Results

### 3.1 Hardness characterization after nanocrystallization

Fig. 1 a) presents the hardness of the CoCrNi and NCoCrNi alloys after HPT processing measured at the flat surfaces of the HPT disks. The radially resolved hardness values are the mean value of 4 measurements from a sample, visualized in reference to the position of the center of the disk. The results show that the hardness values reach a plateau already at ~1mm distance from the center, maintaining this value fairly stable across the radius. As explained, the hardness value and the microstructure are expected to be very homogenous in the areas of subsequent mechanical testing. The alloying of the CoCrNi material with N results in an increase of mean hardness value in the plateau region by 6 % after HPT, from 598 to 631 HV0.1.

Radially-resolved hardness measurements were also performed on the polished cross-sections of the HPT disks in two lines near the surface and at the middle of the cross-section presented in Fig. 1 b). This was done to analyze the hardness homogeneity in the bulk of the disks. The mean hardness values measured on cross-section close to the surface of the disks are almost identical with the mean hardness measured before on the flat surfaces (Fig. 1 a)), showing again ~6% difference between the CoCrNi and the NCoCrNi alloys. The hardness values measured on cross section in the middle of the disks from both alloys show consistently slightly higher (~2%) mean values compared to the sample surface, which is in agreement with results from [28].

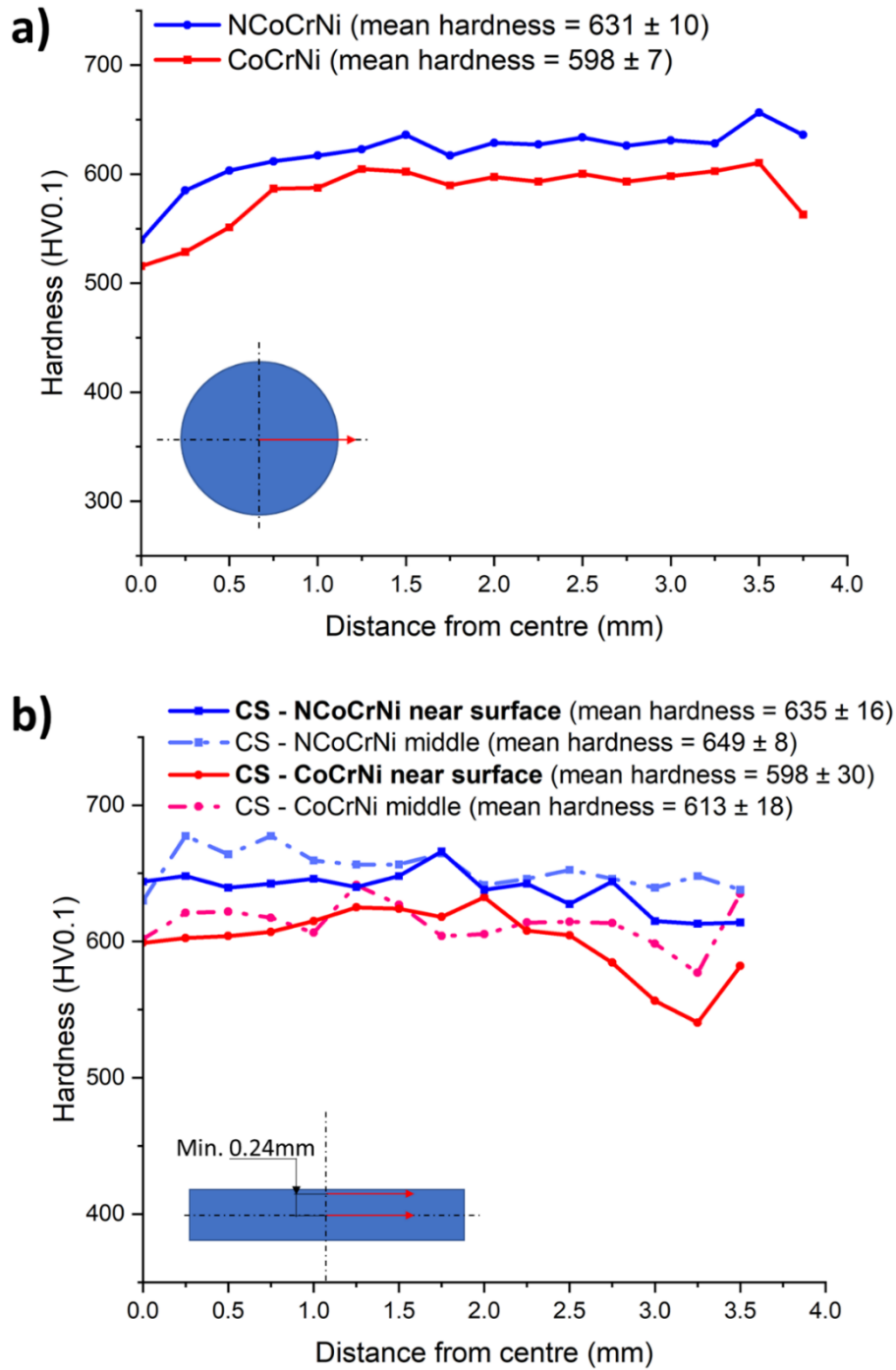

*Fig. 1 Hardness distribution of CoCrNi and NCoCrNi subjected to HPT as a function of the radial position: a) hardness measured at the polished circular face of the 8mm HTP-processed disks; b) hardness measured on polished cross section (CS) of the disk at 2 locations (near the surface and at the central line of the disk CS). The red lines in the sample sketches illustrate the direction of the hardness line measurement.*

### 3.2 Hardness development upon annealing

The hardness evolution after annealing treatments for 1h are displayed in Fig. 2a. The hardness measurements were performed on the circular flat surfaces of the HPT disks, similar to the results from Fig 1a). After an initial slight drop in hardness at 150°C, the mean hardness starts to increase with increasing annealing temperature, reaching a maximum hardness of 681±8 HV0.1 at 450°C for the HPT-processed CoCrNi alloy. The relative increase of the annealed state hardness in comparison to the HPT-processed state is ~14 %, which is slightly lower compared to the value found for the Cantor alloy (21%) processed and thermally annealed under the same conditions [27]. On the other hand, the N-alloyed CoCrNi alloy shows a maximum hardness of 800±9 HV0.1 at 500°C with ~27 % relative increase compared to HPT-processed condition. These results suggest that the strengthening mechanism may be altered by N, based on the larger hardness increase at slightly higher temperature, which will be discussed later in more detail. After the intermediate temperature hardness increase at 450-500°C, the hardness in both alloys rapidly drops to relatively similar values of 212±5 HV0.1 and 224±5 HV0.1, for the CoCrNi and NCoCrNi alloys, respectively annealed at 1000°C.

The results of isothermal annealing at 300°C and 500°C are presented in Fig. 2b. The hardness development at 300 °C follows the same trend for both materials, the only difference being constantly ~9-10% higher values for the NCoCrNi. The hardness at 300°C reach their peak values after 100 h showing a ~10% increase. On the contrary, the hardness changes at 500°C are different compared to annealing at 300°C. While the hardness of the CoCrNi alloys rapidly decreases at 500°C, the hardness of the NCoCrNi alloy remains relatively stable until 80 h of annealing time. The hardness decreases after the longest 120 h annealing time by 37% for CoCrNi and only by 10% for NCoCrNi compared to the HPT deformed starting state.

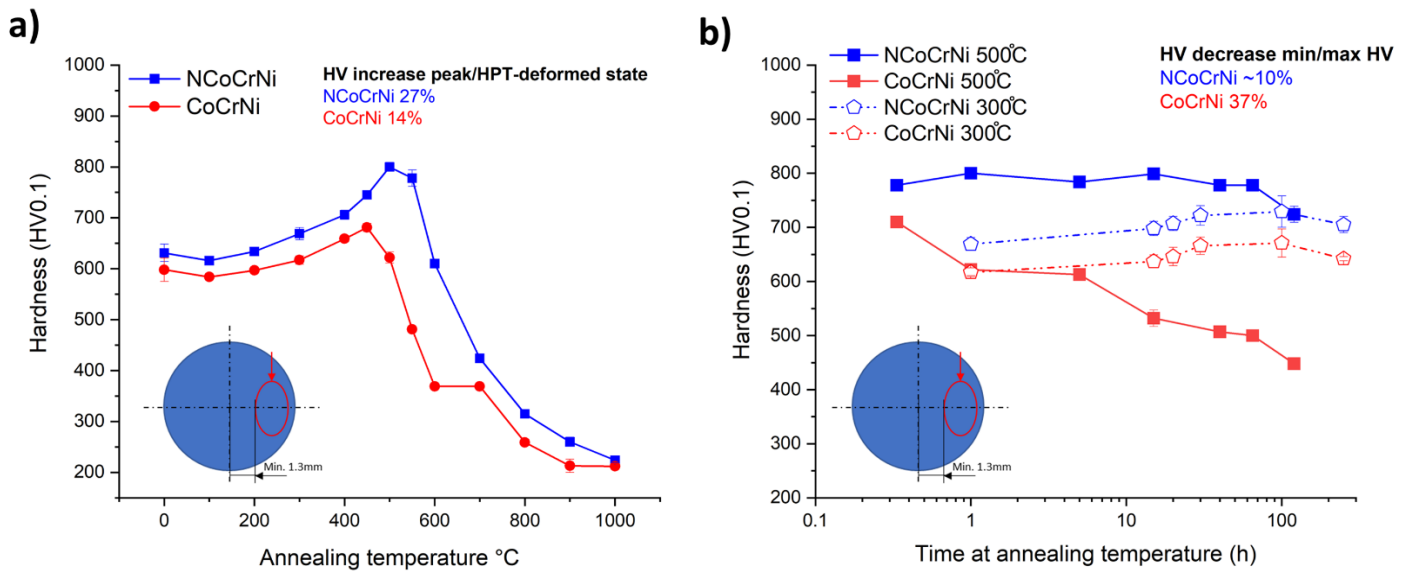

*Fig. 2 Microhardness of the CoCrNi and the NCoCrNi alloys after: a) 1h isochronal annealing at various temperatures; b) isothermal annealing at 300 °C and 500 °C for various times. Typical region where measurements were taken are included in the charts.*

**3.3 Microstructural features of the as-processed states**

Fig. 3 presents the microstructural analysis of the HPT-processed CoCrNi and NCoCrNi alloys. Figs.3 a) and b) present TEM images of the highly deformed, nano-grained FCC-microstructures, respectively. Both materials show very similar microstructural features using classical bright-field imaging. The subsequent TKD analysis presented in Figs. 3c) and d) revealed a difference in mean grain size. While the CoCrNi alloy shows a mean value of 74±30 nm, the NCoCrNi alloy shows a ~42% decrease to 43±30 nm. Therefore, it seems that the nitrogen addition reduces the grain size substantially. It should be noted that both TKD orientation maps show a large fraction of unindexed, black areas, which is ~30% higher in NCoCrNi compared to CoCrNi. The inability to index certain areas is in the case of the TKD analysis connected to the small grain-size and internal stresses. The fact that the NCoCrNi map has a larger fraction of unindexed area is consistent with its observed smaller grain size, as compared to the CoCrNi. A weak shear texture (not shown), typical for FCC HPT processed metals [33][34] was found. Grain size and texture analysis was performed on 2-3 TKD datasets in order to get better statistics.

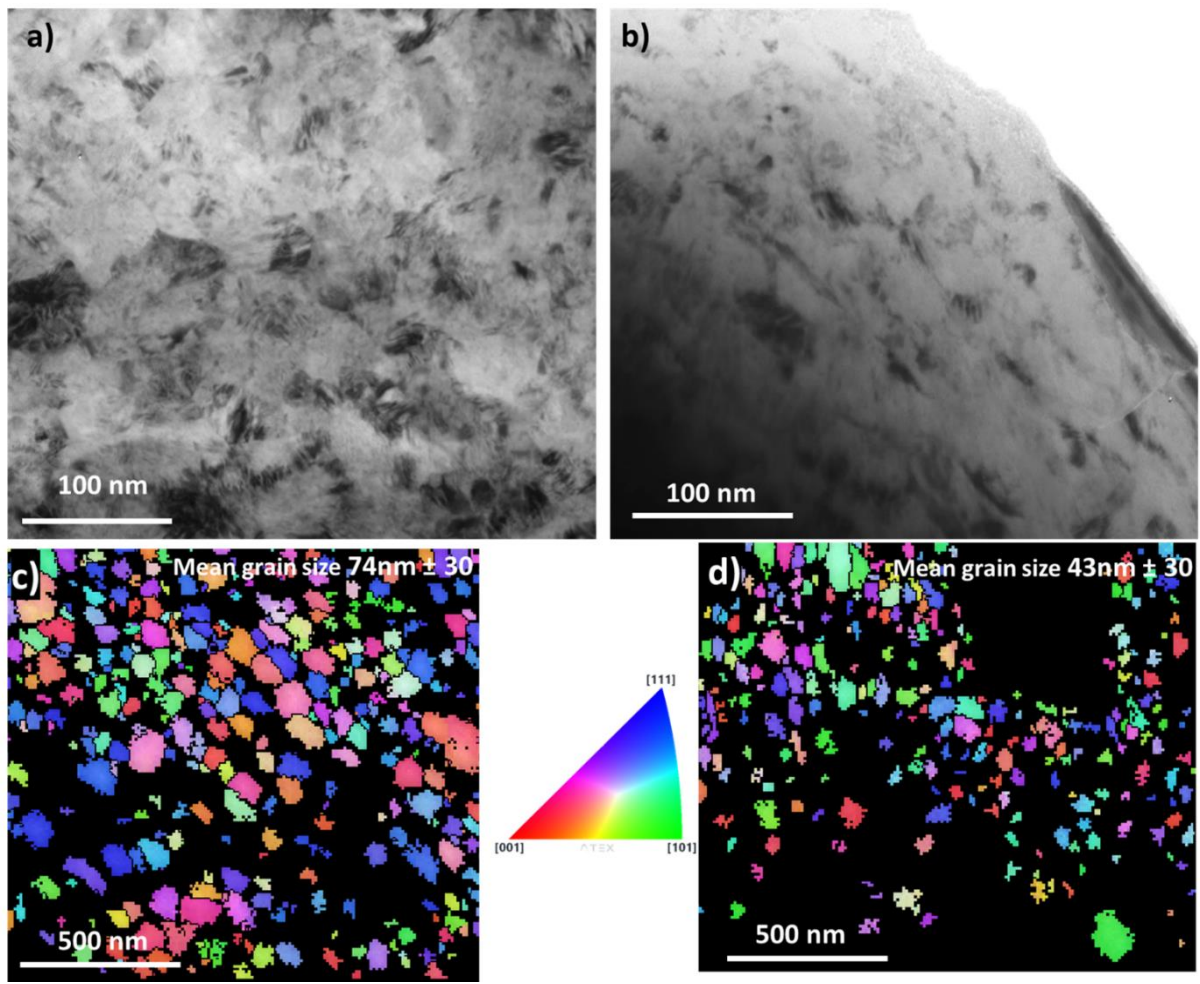

*Fig. 3 Microstructural analysis of CoCrNi (a,c) NCoCrNi (b,d) after HPT: TEM images showing representative microstructures with nano-sized grains of the FCC matrix phase of CoCrNi a) and NCoCrNi b); TKD orientation*

maps of CoCrNi c) and NCoCrNi d) with reference direction parallel to the direction of applied force during HPT; Note that the black region in the TKD orientation maps are unindexed areas due to insufficient signal.

### 3.4 Bending properties

The results of the micro-cantilever bending experiments are summarized in Fig. 4, while the statistical values of characteristic properties are presented in Table 1. The materials show very high strength levels reaching a proof stress $\sigma_{bp0.2}$ of 2421±25 MPa and a maximum stress $\sigma_{bm}$ of 3655±9 MPa for the CoCrNi alloy. The NCoCrNi alloy shows a higher overall strength, with proof stress $\sigma_{bp0.2}$ and maximum stress $\sigma_{bm}$ being 14.6% and 8.8% higher compared to the CoCrNi alloy, respectively. Note that none of the cantilevers from the HPT-processed alloy condition fractured in a brittle manner, but rather deformed plastically until reaching the maximum displacement allowed by the used setup. This reveals that both materials maintain distinct ductility levels even in the nano-crystalline state. The materials also show quite significant strain hardening of 823 MPa and 803 MPa for the CoCrNi alloy and the NCoCrNi alloy, respectively, calculated as difference between proof stress $\sigma_{bp0.2}$ and maximum stress $\sigma_{bm}$.

In order to test larger material volumes additional bending tests were performed on miniaturized 3-point bending samples with a characteristic width W ~ 1 mm. The results are summarized in Fig. 5 and Table 2. Specimens from the HPT-processed state and after annealing at 500°C for 1 h were tested. The results of the HPT-processed samples show the same trend of N-induced strength increase as found for micro-cantilever testing, while the overall amounts of stress are reduced by ~45-60%. This drastic difference will be discussed later in more detail. Nevertheless, both material states again show high strength levels. The HPT-processed CoCrNi alloy has a bending proof stress $\sigma_{3pb0.2}$ of 1044±37 MPa and maximum stress $\sigma_{3pbm}$ of 1528±17 MPa. The N alloyed HPT-processed NCoCrNi alloy, compared to CoCrNi, exhibits a 41% and 33% increase in proof stress $\sigma_{3pb0.2}$ and maximum stress $\sigma_{pbm}$, respectively. These differences measured for miniaturized 3-point bending samples are much larger compared to the differences measured with micro-cantilevers (14.6 % and 8.8 %), but the overall trend is identical. As opposed to the micro-cantilever, 3-point bending beams from both materials did all break during testing, albeit showing significant plasticity with a fracture strain >0.05 in all cases. Interestingly, the mean fracture strain of the NCoCrNi alloy (~0.1) is ~40% larger than the value for CoCrNi (~0.07), despite showing significantly higher proof and maximum stress. Both HPT processed materials also show considerable strain hardening capacity in terms of the difference between $\sigma_{3pb0.2}$ and $\sigma_{3pbm}$ with NCoCrNi showing a 17% higher value of 567 MPa. Therefore, the ductility of the nanocrystalline state shown before in the micro-scale experiment was retained even in the meso-scale tests. On the other hand, the CoCrNi and NCoCrNi samples subjected to intermediate temperature annealing at 500°C display brittle fracture behavior. Contrary to expectations, the fracture stress of 1695 MPa for the 500 °C annealed CoCrNi alloy is 45% higher than the fracture stress for the NCoCrNi, which will be discussed later.

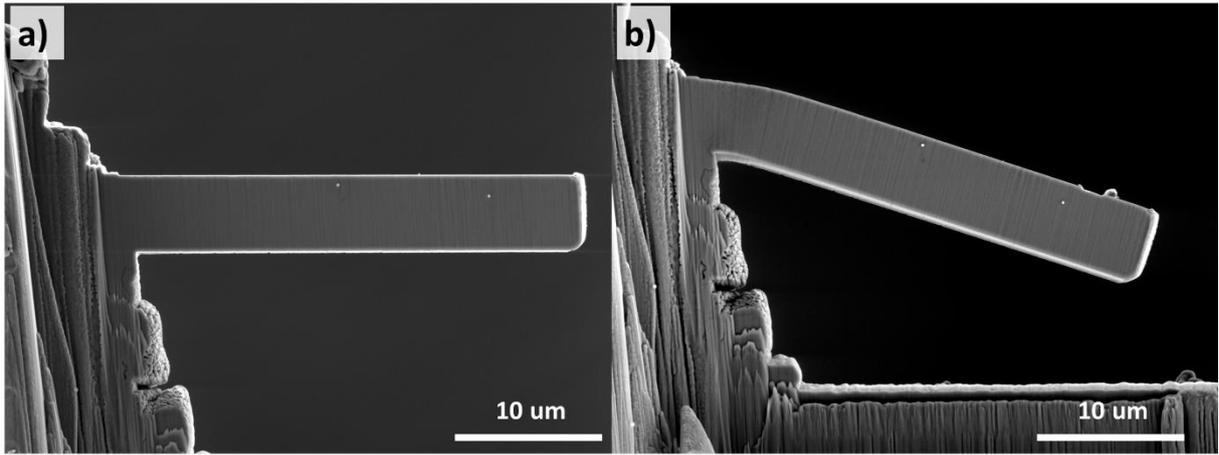
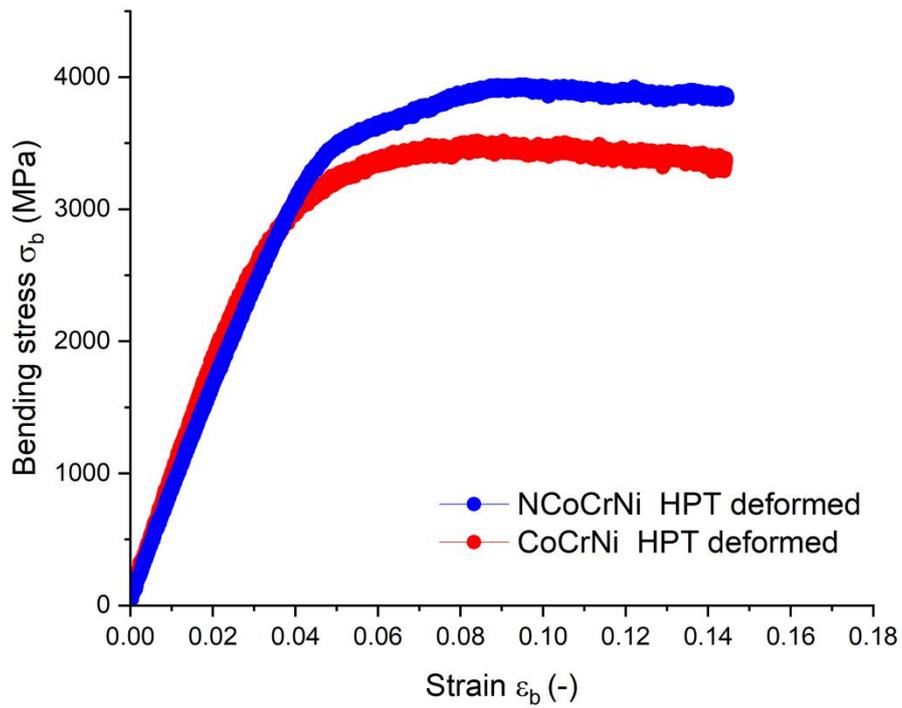

*Fig.4 Results from the micro-cantilever bending experiments in the HPT-processed state. Representative cantilever from the NCoCrNi alloy: a) before bending experiment; b) after experiment; c) Representative bending stress-strain curves. None of the cantilevers fractured during testing.*

Table 1 Bending properties of the tested micro-cantilevers in HPT processed state. The mean values from 3-4 cantilevers are presented.

|  | $\sigma_{bp0.2}$ (0.2% strain plastic stress) (MPa) | $\sigma_{bm}$ (maximum stress) (MPa) | $\sigma_{bm} - \sigma_{bp0.2}$ (strain hardening) (MPa) | $E_{b}$ (bending modulus)(GPa) |
|---|---|---|---|---|
| CoCrNi | 2421±25 | 3655±91 | 823 | 114±16 |
| NCoCrNi | 2775±151 | 3979±83 | 803 | 117±4 |

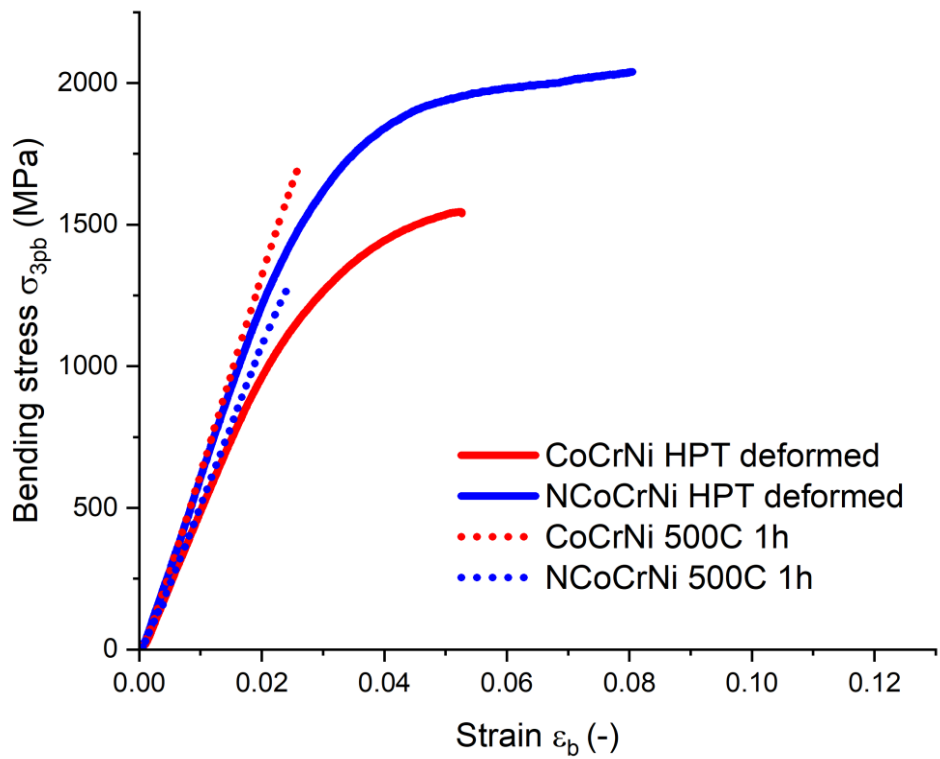

Fig. 5 Representative bending stress-strain curves of miniaturized 3-point bending samples in the HPT processed state compared to 500°C 1 h annealed state.

Table 2 Bending properties of miniaturized 3-point bend samples in HPT processed state compared to 500°C 1 h annealed state. The mean values from 2 beams are presented, except for CoCrNi 500°C 1 h state where only one beam was tested.

| | $\sigma_{3pb0.2}$ (0.2% strain plastic stress) MPa | $\sigma_{3pbm}$ (maximum stress) MPa | $\sigma_{3pbm} - \sigma_{3pb0.2}$ (strain hardening) MPa | Fracture strain |
|---|---|---|---|---|
| CoCrNi after HPT | 1044±37 | 1528±17 | 484 | 0.055±0.0025 |
| NCoCrNi after HPT | 1469 ±26 | 2036±26 | 567 | 0.081±0.019 |
| CoCrNi 500°C 1h | 1695 | | - | 0.026 |
| NCoCrNi 500°C 1h | 1140±16 | | - | 0.023±0.0005 |

The fracture surfaces of the fractured miniaturized 3-point bending specimens from Fig. 5 are displayed in Fig. 6. All fracture surfaces are composed of dimples with non-metallic particles inside the dimple cusps, revealed to be mostly Cr-rich oxide inclusions by EDS analysis. The HPT-processed CoCrNi alloy in Fig. 6 a) and show smaller dimples compared to the NCoCrNi material in Fig. 6 b) and slightly less prominent occurrence of rounded inclusions.

The fracture surfaces after 500°C 1 h annealing display void sheets composed of extremely fine dimples. Such fine voids organized in sheets without inclusion particles may have formed on grain boundary triple junctions in later stages of fracture, due to the triaxial stress states occurring in areas around the larger dimples that formed first from larger inclusions [35]. The fracture surface of annealed NCoCrNi is considerably more heterogeneous compared to the annealed CoCrNi. Except for the rounded Cr-rich oxide inclusions, cuboidal inclusions were detected on the NCoCrNi fracture surface, visible in the inset of Fig. 6 d). They correspond to Ti-Cr-rich cuboidal carbide inclusions.

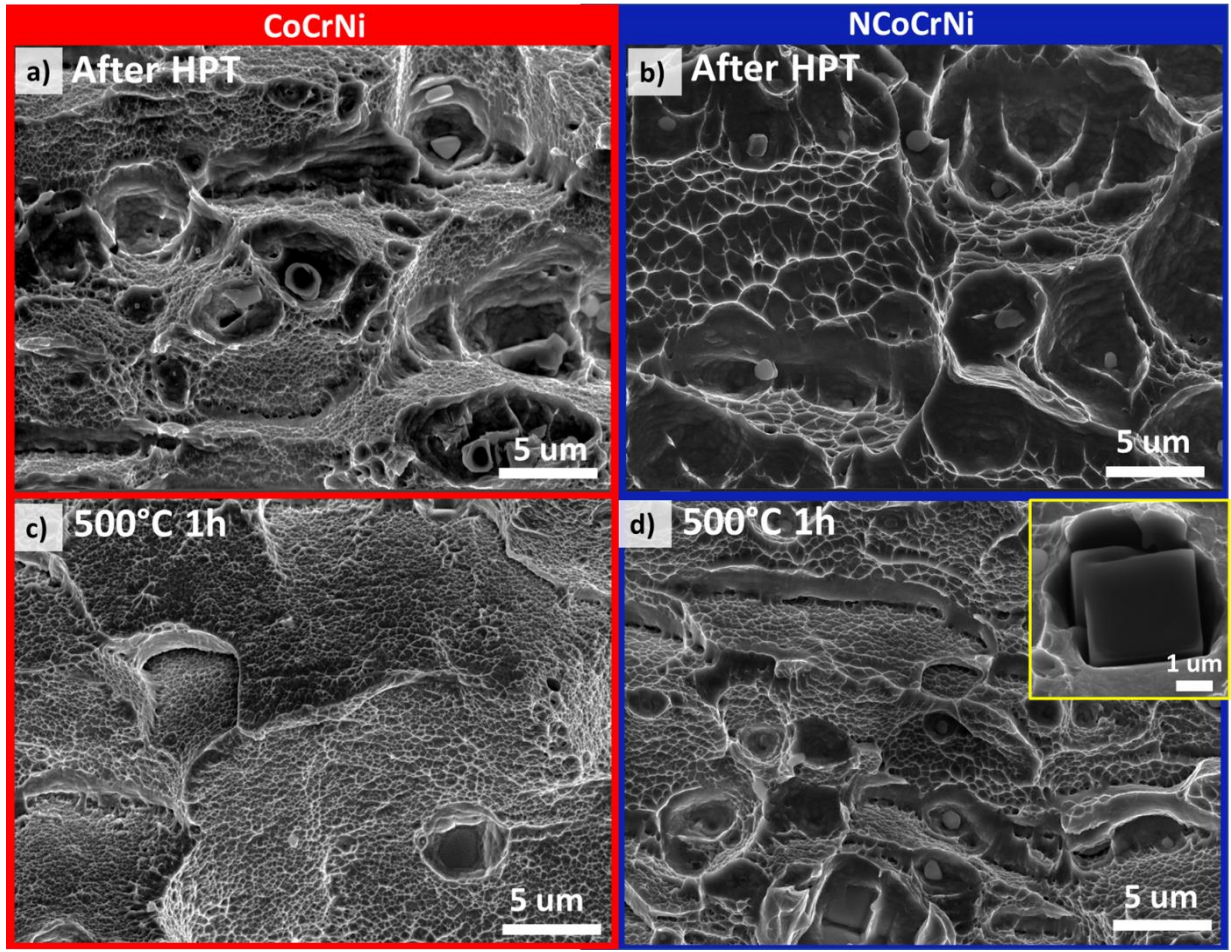

*Fig. 6 The Fracture surfaces of the representative miniaturized 3-point bending specimens after HPT from CoCrNi (a) and NCoCrNi (b); and after 500°C 1h annealing from CoCrNi (c) and NCoCrNi (d) alloys. Fracture surfaces with varying size of dimples are visible in all fracture surfaces. Larger dimples are observed after HPT while very fine dimples organized in void sheets are preset in 500°C 1h annealed state.*

**3.5 Microstructural development after 1h isochronal annealing**

Grain growth was investigated after isochronal annealing for 1 h annealing at various temperatures. The microstructures after annealing at temperatures below 450°C did not yield changes in microstructural features that could be adequately resolved by SEM, even though there was a substantial increase of the measured hardness. The microstructures after HPT-processing compared to microstructures after annealing at 500 °C for 1 h are shown in Fig. 7. While the CoCrNi alloy after 500°C anneal in Fig. 7 c) already contains areas with partially recrystallized grains, the microstructure of NCoCrNi after 500 °C anneal in Fig. 7 d) does not show any significant changes compared to the HPT-processed state.

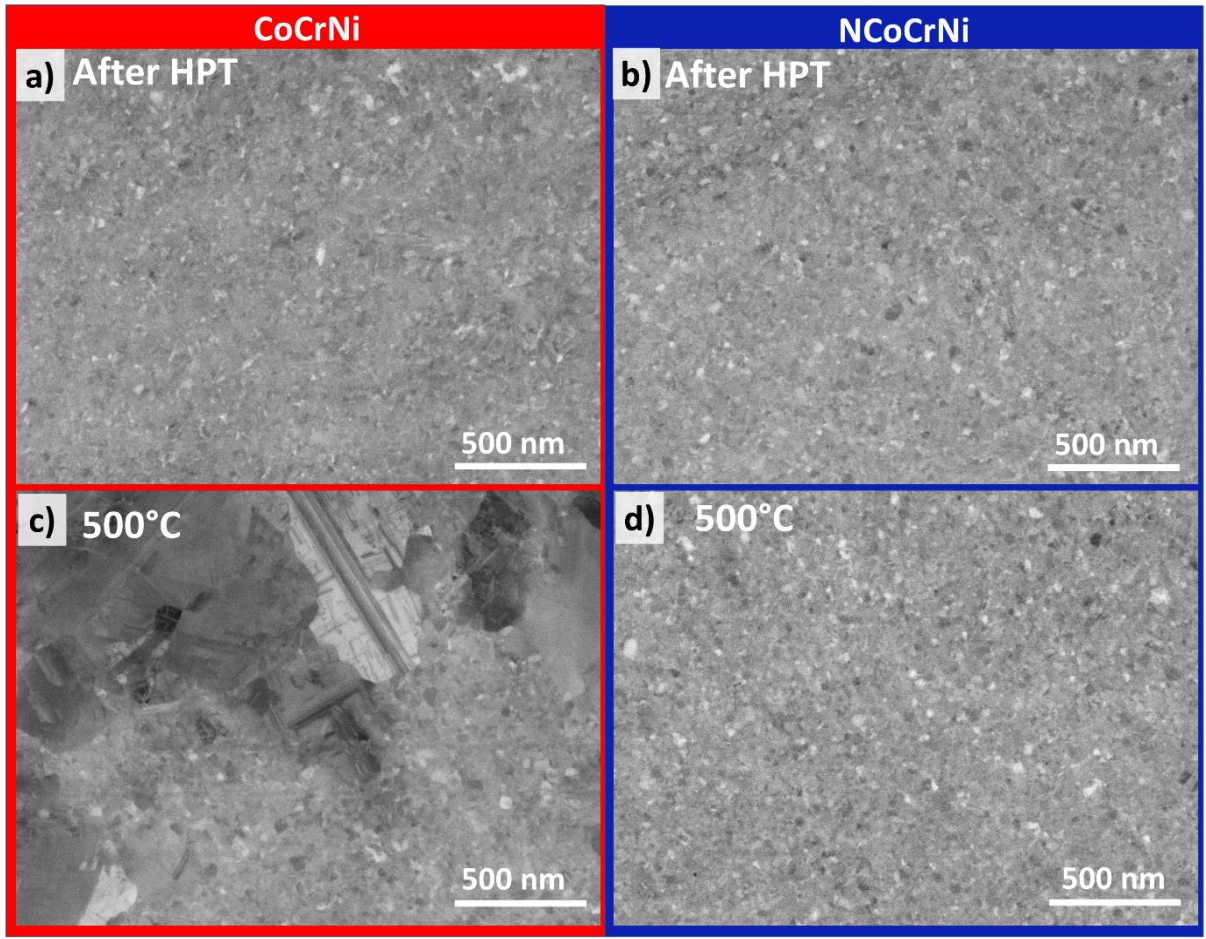

*Fig. 7 Representative microstructures of – CoCrNi (a, b) and NCoCrNi (c, d) materials after HPT-processing (a, c); and annealing for 1h at: 500 °C (c, d).*

Some of the representative SEM microstructures after annealing at temperatures 550°C and above are displayed in Fig. 8 and mean grain sizes from EBSD analysis are summarized in Table 3. The most evident feature of the microstructures after 550 °C annealing in Figs. 8 a, b and 600°C annealing in Figs. 8 c, d is the clear difference in the grain size between CoCrNi and NCoCrNi. After 550°C annealing, the recrystallized grains in the CoCrNi alloy are already clearly defined, with a visible dislocation substructure inside. Simultaneously, the NCoCrNi alloy after 550°C still shows limited grain growth in comparison to the as-deformed state. At 600°C, the grain size increases slightly and was measurable by EBSD with 1.14 µm for CoCrNi and a 65% lower value of 0.69 µm for NCoCrNi. While the grain size after 600°C annealing in CoCrNi is relatively homogenous with evident dislocations and stacking faults; green arrow in Fig. 8 c), the NCoCrNi grains have a bimodal distribution.

After annealing at 700°C, a secondary Cr and N-rich darker phase was observed in NCoCrNi, denoted by yellow arrows in Fig. 8 f). The phase resembles non-metallic inclusions, but its number density is higher. This phase was identified by the EDS to be rich in Cr and N, pertaining to Cr-nitride. Due to the high concentration of light nitrogen, the phase gives a similar dark contrast comparable to mostly oxygen-based inclusions in the FCC matrix with higher atomic mass (Co, Cr, Ni).

From 800 °C annealing temperature up to a maximum of 1000 °C (Figs. 8 g, h), there are no traces of any secondary phases visible in both materials that formed from the intentional alloying elements. The dark spots in the micrographs correspond to non-metallic inclusions rich in oxygen, based on EDS analysis. The differences between the grain size of CoCrNi and NCoCrNi are between 53-75% up to 800 °C annealing temperature. At 900 °C, the difference in grain size between the materials drops to 38%. The fully recrystallized single-phase FCC microstructures of both materials visible after 1000 °C anneal displayed in Figs. 8 e) and f) show only a 18% difference in grain size with mean values of ~33 µm and ~28 µm for CoCrNi and NCoCrNi. Σ3 annealing twin boundaries compose a large fraction (up to 40-50%) of the grain boundaries, which is typical for low stacking fault energy materials such as CoCrNi-based alloys [36].

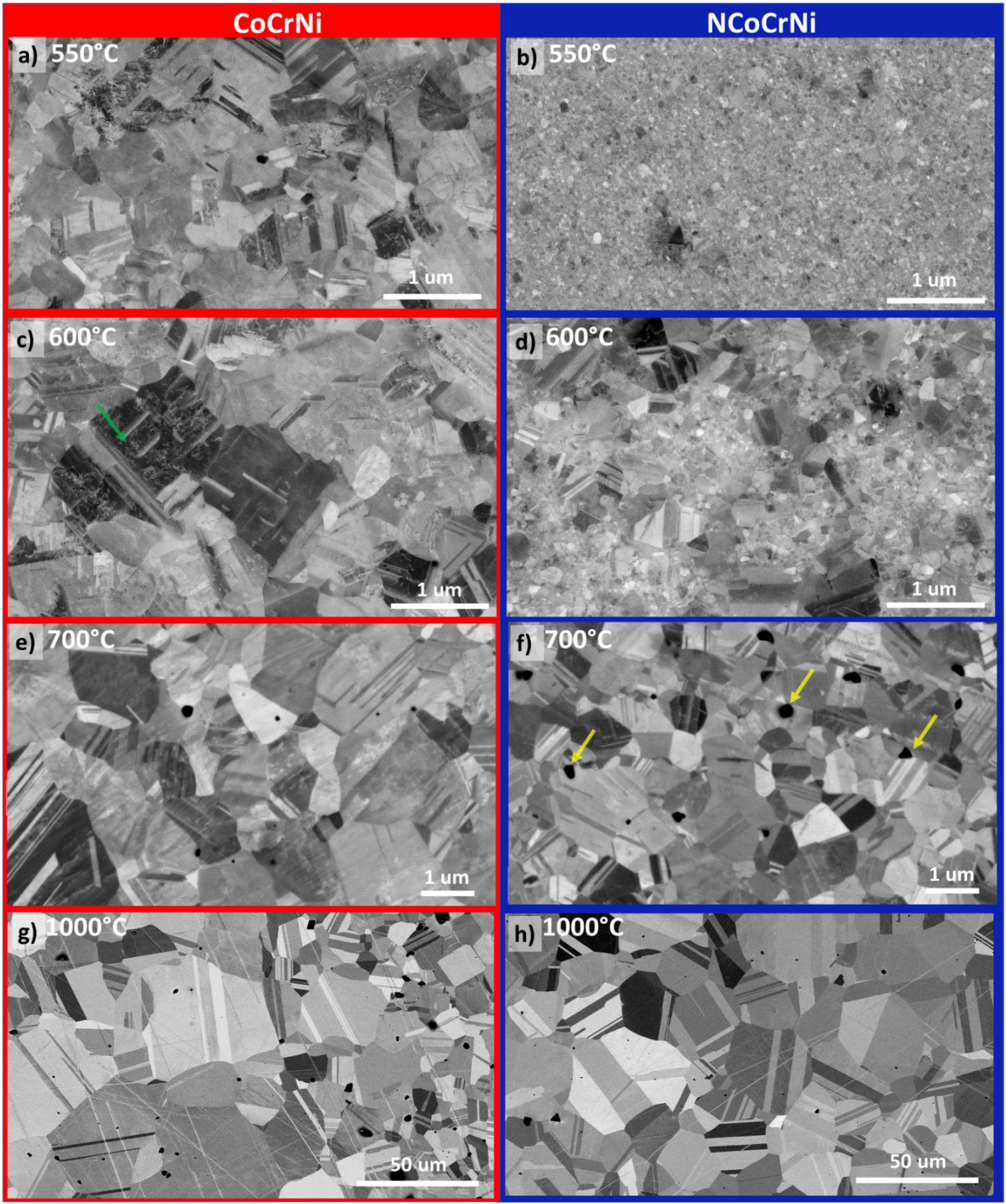

*Fig. 8 Representative microstructures of – CoCrNi (a, c, e, g) and NCoCrNi (b, d, f, h) materials after HPT and isochronal annealing for 1h at: 550 °C (a, b); 600°C (c, d); 700°C (e, f); 1000°C (g, h). The dark spots in all materials correspond to conventional contamination by non-metallic inclusions (oxide based) except for the points in f) denoted by yellow arrow that represent Cr, N-rich phase.*

*Table 3 Mean grain size of the specimens estimated by EBSD experiments after annealing for 1 h.*

| Annealing Temperature (°C) | Mean grain size (µm), CoCrNi | Mean grain size (µm), NCoCrNi | Relative difference % |
| --- | --- | --- | --- |
| 600 | 1.14 | 0.69 | 65 |
| 700 | 1.92 | 1.25 | 53 |
| 800 | 7.39 | 4.22 | 75 |
| 900 | 20.3 | 14.7 | 38 |
| 1000 | 33.4 | 28.2 | 18 |

**3.6 Microstructural development after isochronal annealing at 500°C for 120 h**

The long-term annealing at 500°C for 120 h resulted in significant differences in the measured hardness and therefore further in-depth microstructural characterizations were carried out. The STEM micrograph, TKD orientation map and grain size distribution of CoCrNi in Fig. 9 a), c) and e) show a relatively homogenous grain size with a mean value of 1.34 um. Compared to CoCrNi, the grain size distribution of NCoCrNi in Fig. 9 b), d) and f) exhibits larger area fraction of smaller grains, with a 62% smaller mean value of 0.52 um. Further investigations of the NCoCrNi microstructure by TEM EDS in Figs. 10 a)-c) reveal, that the alloys contain an additional phase of hexagonal $Cr_2N$ grains between the FCC matrix. The phase is present in the NCoCrNi material on the boundaries of the regions with distinctively different grain sizes and inside the nano-grain sized regions. This phase was also observed in the microstructure by conventional SEM-EDS and corresponds to the dark phase observed during 1 h annealing experiments at 700°C (cf. previous section Fig. 8 f).This secondary phase is most likely the reason for the observed bimodal microstructure found for annealing treatments 600 °C for 1 h (see Fig.8 d) and 500 °C for 120 h (see Fig.9 f).

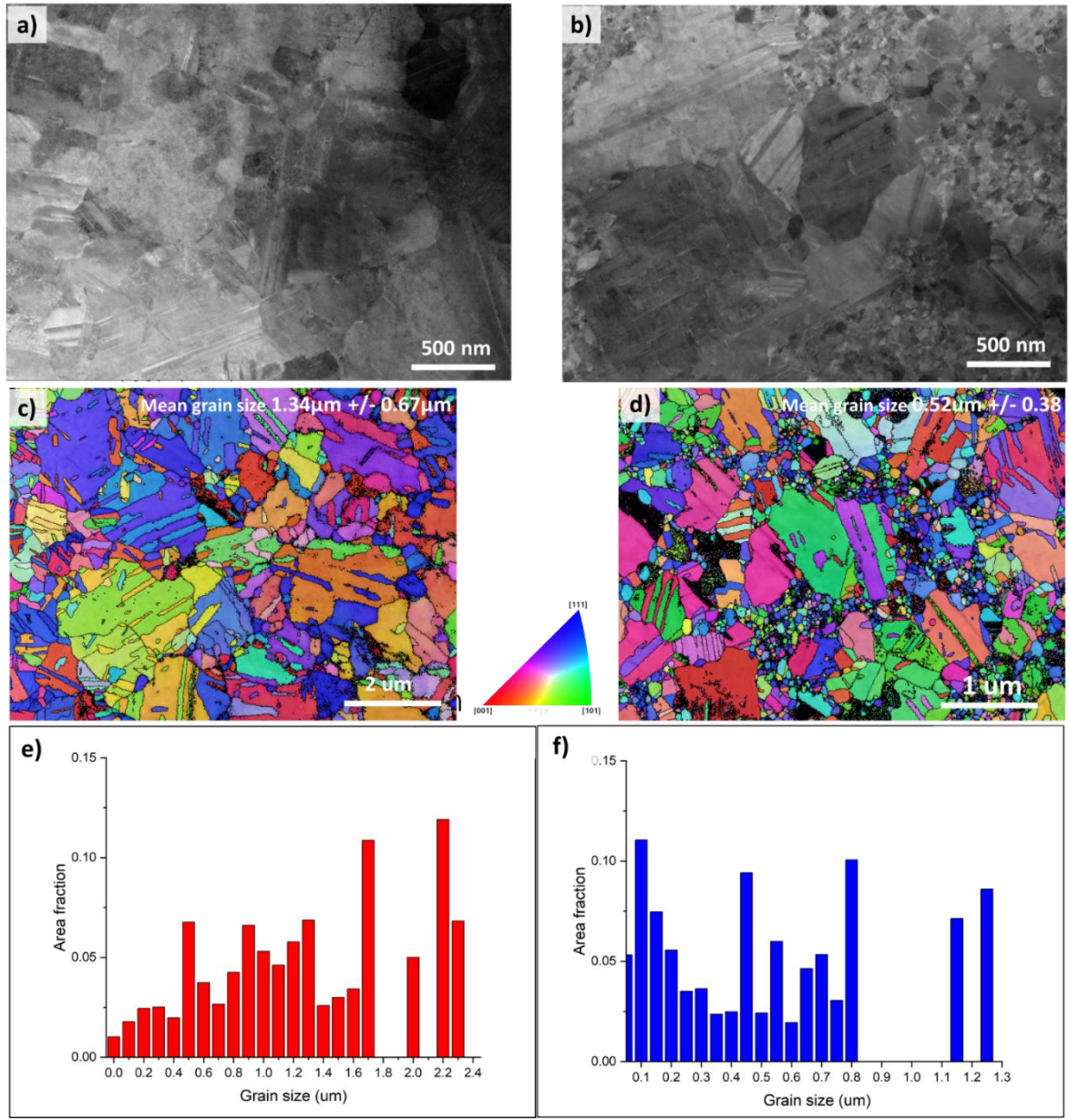

*Fig. 9 Representative microstructures of the CoCrNi (a, c, e) and NCoCrNi (b, d, f) materials after HPT and 500 °C annealing for 120 h. a-b) STEM images (SEM based); c-d) TKD orientation maps with visualization of the orientation distribution in axial HPT-direction; e-f) grain size – area fraction histograms. Note that the black region in the TKD orientation maps are unindexed areas due to insufficient signal. Note that the scale bars and sizes of the maps in b) and d) are different.*

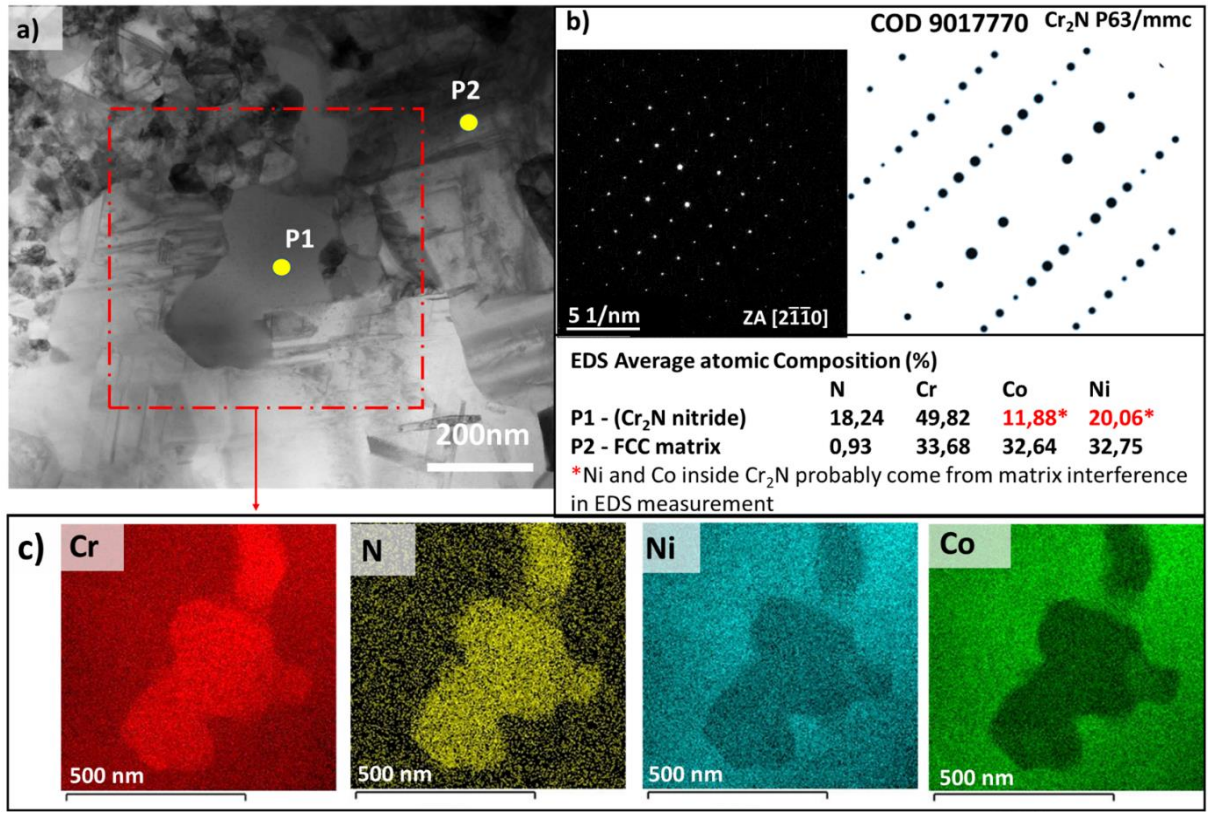

*Fig. 10 The results of TEM characterization of the nitrogen-alloyed NCoCrNi alloy annealed at 500 °C 120h. a) bright-field STEM image (TEM-based) from the area of interest at the boundary of two regions with distinctively different grain size; b) selected area diffraction pattern from point P1 and simulated diffraction pattern of the $Cr_2N$ and EDS elemental compositions of points P1 (precipitate) and P2 (FCC matrix); c) EDS elemental maps from the region marked by red rectangle.*

## 4. Discussion

### 4.1 Influence of nitrogen on mechanical properties and thermal stability

While it was shown that N alloying increases the hardness by increasing solid solution strengthening [9], the results of this study also reveal that a more pronounced grain refinement during severe plastic deformation is achievable by the addition of N. These two combined phenomena of nitrogen-supported grain refinement and solid solution strengthening are responsible for ~6% increase in hardness and 15-33% increase in maximum strength after HPT, compared to N-free CoCrNi material. Despite the achieved nanocrystalline state with grain sizes on the order of tens on nanometers, both HPT-processed materials still show considerable strain hardening capacity and ductility, where micro-cantilevers and meso-sized 3-point bending beams can withstand a substantial amount of bending strain without failure. The plasticity after HPT is also apparent from the observation of ductile dimples on the fracture surfaces of both alloys, displaying a wide range of diameters.

The presence of alloying elements in low concentrations, or even trace levels of impurities generally increase the magnitude of grain refinement after SPD processing [24]. The effects of interstitial-induced grain refinement after SPD were already observed for dissolved C in pure nickel [35], hyper-eutectoid steels [22] or Fe-C coatings [37]. In heavily-deformed alloys, C segregation at sub-grain boundaries formed during SPD suppresses dynamic recovery [38], resulting in grain refinement. The effect of N is probably less pronounced compared to C, due to the lower grain boundary segregation tendency visible from APT measurements on CoCrNi [39,40] or BCC steels [41][42]. This tendency can be rationalized by a larger decrease of the grain boundary free energy $G_{GB}$ with C segregation in comparison to N segregation [42][43]. This results in a higher stability (lower mobility) of grain boundaries with segregated C, compared to segregated N. However, due to the higher tendency of C to segregate at grain boundaries and comparable affinity towards Cr compared to N [44], local conditions favor formation of stable Cr carbides in the C-CoCrNi system. The carbides in CCoCrNi with 0.5 at% of C form after annealing below 800°C and remain in the microstructure until solution annealing at 1280°C [45], as compared to nitrides in the NCoCrNi alloy that are observed at lower temperatures and readily dissolve after annealing at ~800°C.

The effect of N on the hardness after the 150-400 °C annealing treatment is similar to that directly after HPT. The results of the annealing treatments at temperatures up to 300-400 °C show identical trends for both materials, with steadily increasing hardness values. N-alloyed CoCrNi show constantly 10% higher hardness values compared to CoCrNi, similar to the difference after HPT. Therefore, the same strengthening mechanisms of solid solution and grain boundary strengthening operate analogous as for the HPT processed state, with extra strengthening originating from recovery of dislocation substructures at grain boundaries causing reduction of dislocation nucleation sites, as described in previous work [46].

At annealing temperatures in the range 400°C - 600°C, the differences caused by N-alloying are more pronounced. The N-alloyed CoCrNi shows maximum increase of hardness compared to the HPT-deformed state by ~28% at temperature of 500°C, while the unalloyed CoCrNi shows the largest increase by ~14% at 450°C – compared to ~20% increase from previous investigations on the interstitial-free Cantor alloy and CoCrNi alloys [27,47]. Interestingly, the hardness of N-alloyed CoCrNi is much more stable during isothermal annealing at 500°C, which can be associated with less pronounced grain coarsening, compared to CoCrNi. However, the 3-point bending tests after 500°C annealing reveal a complete loss of macroscopic ductility or strain hardening for both materials, contrary to the HPT processed state. Interestingly, the observation of extremely fine voids on the fracture surfaces is indicative of local micro-plasticity during fracture, similar to results on 316L steel treated in a same way [48]. However, this local micro-plasticity has no effect on the resulting macroscopic brittle fracture behavior. The latter is caused by plasticity confinement to a very small area due to strain softening after initial onset of plasticity and micro-void formation, similar to HPT-processed and annealed 316L steel [48].

Despite a higher hardness after 500°C annealing, the 3-point fracture stress of NCoCrNi is lower than for the CoCrNi material. This result can be rationalized on the basis of large sensitivity to microstructural defects and a large fracture stress scatter for materials exhibiting a completely brittle fracture [49]. The similar loss of plasticity after annealing at 150-500°C temperature was already observed for the CoCrNi alloy [29] and some other metals such as stainless steels, nickel or aluminum [50][51][52][24]. The hardening effect, together with embrittlement is attributed to either grain boundary segregation and consequential reduction of the mobility of grain boundary dislocations or annihilation of mobile dislocations at grain boundaries due to thermally induced relaxation [46,53][54]. In the case of NCoCrNi, the formation of nitrides observed after 500°C annealing may also contribute, to some extent, to the hardening effect, as discussed in section 4.2.

At annealing temperatures from 700°C up to 900°C, the differences in hardness caused by the N-alloying again drop to ~10%, until the difference almost disappear after 1000°C annealing. The differences in grain sizes at these temperatures follow a similar trend as the hardness values. The decreasing differences in the temperature range of 700°C - 1000°C compared to 400°C - 600°C corresponds to the disappearance of secondary nitride particles that seem to dissolve in the FCC matrix at annealing temperatures at and above 800°C. The disappearance of nitrides at temperatures 800°C and higher, is supported by previous results on identical N-alloyed CoCrNi material, where no secondary particles were observed [40]. It should be noted that even relatively coarse nitride particles are observed at 700°C annealed NCoCrNi, with a size on the order of the grain size and <0.5% fraction, which are not expected to produce significant strengthening.

**4.2 The formation of nitrides**

The observation of nitrides shows that second phase(s) will precipitate from the FCC matrix after exposures at intermediate temperatures, similar as in the case of other N-alloyed systems such as steels [55], Ni [56] and Co-alloys [57,58]. The rate of their formation is probably accelerated by previous SPD due to stored mechanical energy, decreasing the activation barrier for thermodynamic processes such as precipitation. In the case of CoCrNi, there seems to be a temperature range where secondary nitride particles form during annealing after HPT processing. Their presence results in a microstructure with two regions differentiated by distinctively different grain diameters, as observed in NCoCrNi after annealing at 500-600°C (see Fig. 8d). This is a consequence of grain boundary pinning by nitrides. The regions with very small grains contain nano-sized nitrides at the grain boundaries or larger nitrides at the boundaries of the fine-grained zone. The regions with large FCC grain diameters do not contain such nitrides. The nitrides, however, dissolve in the matrix after heating to temperatures at and above ~800°C.

**4.3 Mechanical size effect**

Relatively large differences between the micro-cantilever samples and meso-scale 3-point bending samples were recorded here, *i.e.* the HPT processed CoCrNi alloy show a 58% reduction in the maximum stress measured by 3-point bending (W~1mm), compared to the maximum stress of micro-cantilevers (W~5 um). The micro-cantilevers also strain harden 60% more compared to the meso-scale bending samples.

One approach to understand these differences may be attributed to a mechanical size effect [59][60], influencing also the extent of strain hardening [61]. The size dependent stress $\sigma_s$ can be written as [62][63,64]:

$$\sigma_s = \sigma_{ref}\left(\epsilon^2 + l^\beta \chi^\beta\right)^{n/2} \tag{6}$$

Here, $\sigma_{ref}$ is a constant term in MPa depending on the material, similar to the strength coefficient *K* in Hollomon's power law [65] and the constant *n* is the strain-hardening coefficient, both for macroscopic samples. The variable *ε* is the effective strain with *β=1* [62,66]. The parameter χ corresponds to the occurring strain gradient, strongly depending on the sample size. The characteristic length scale *l* is approximated by the grain size. Choosing typical values based on results obtained by meso-scale testing, i.e. *n*=0.2, $\sigma_{ref}$= 2000 MPa, strain *ε* = 0.1 and *l* (grain size) = 70 nm, the difference in stress between samples with W=1 mm and W=5 um is predicted to be ~1%. Even for W = *l* = 5 um, only 26% difference in stress is predicted. Therefore, the differences in the stress cannot be fully explained by the occurrence of this size effect. Other reasons for the difference that should be considered are:

- slightly different states of stress and strain in the micro-cantilevers, compared to the state presumed for the basic calculations assuming small deformations (equation 1 and 2).

- The specimen orientation in SPD-processed materials is known to be a possible source for mechanical differences [67]. In the present study, the loading axis of the small specimens was the radial HPT-direction, where the large ones were tested in axial direction. This could lead to a difference based on texture which is however known to be weak. Additionally, the axial homogeneity in the hardness distribution across the thickness of the samples with somewhat lower hardness values for the margins (~2%) [28] could contribute to the lower yield bending strength for the larger test specimens.

## 5. Conclusions

The bending strength, hardness and microstructure of the CoCrNi medium-entropy alloy (MEA) and its variant with 0.5 at. % N were investigated after high pressure torsion (HPT) and isothermal or isochronal annealing treatments at various temperatures and times. The differences in the behavior that were observed were analyzed and the results can be summarized as follows:

- Alloying of the CoCrNi alloy with N resultes in a ~40% more pronounced grain refinement after HPT and ~10-15% increased hardness and bending strength, mostly due to solid solution strengthening.
- Both investigated alloys retain considerable strain hardening capacity in the nanocrystalline state, HPT processed state with homogenous microstructure and hardness,
- Annealing at intermediate temperatures of 300-500°C for 1 h induces a 20-41% increase in hardness and strength, more pronounced in case of NCoCrNi, showing also better thermal stability. The difference can be explained by the lower HPT-grain size in the N-alloyed material combined with grain boundary pinning induced by the N-addition. For the long-term stability, also $Cr_2N$ precipitation in NCoCrNi plays an important role.
- The largest increase in hardness compared to the HPT-processed state is found after 1 h annealing at 450°C for CoCrNi (14%) and after 500°C annealing for NCoCrNi (29%). This is accompanied by a complete loss of ductility.
- After annealing at temperatures above 800°C, up to 1000°C, both CoCrNi and NCoCrNi exhibit single-phase FCC microstructures, while the differences in hardness and grain size continuously decrease with increasing annealing temperature.

To summarize the findings of the study, combining high substitutional and interstitial strengthening effects in plastically deformable solid solutions leads to stabilization of the nano-grained microstructure obtained by severe plastic deformation. A high-strength material with properties tunable by simple annealing treatment can be produced. The future optimization of interstitial elemental species may lead to potentially further enhanced strength properties without deteriorating the ductility of the alloy.

**Acknowledgements:**

This work was supported by the project from the European Union's Horizon 2020 Research and Innovation Programme under the Marie Skłodowska-Curie Grant Agreement N°101062549. The support of Gabriele Felber, Herwig Felber, Peter Kutlesa, Marco Reiter, Robin Neubauer, Yong Huang and Philip Hobenreich is further acknowledged for technical support and stimulating discussions.

**Data availability statement:**

The raw data required to reproduce a part of these findings are available to download from: https://www.oeaw.ac.at/esi/research/deformation-fatigue-and-fracture/recycling-oriented-alloy-design-for-next-generation-of-sustainable-metallic-materials. The entirety of the raw/processed data required to fully reproduce these findings cannot be shared at this time due to technical limitations. Data can be shared upon request after contacting the corresponding author.